\documentclass[%
 reprint,
superscriptaddress,
 amsmath,amssymb,
aps,
floatfix,
]{revtex4-2}

\usepackage{graphicx}
\usepackage{dcolumn}
\usepackage{bm}
\usepackage{braket}%
\usepackage{here}
\usepackage[colorlinks=true, allcolors=blue]{hyperref}
\usepackage{color}

\begin{document}

\preprint{APS/123-QED}

\title{Description of inclusive $(d,d^{\prime}x)$ reaction with the semiclassical distorted wave model}

\author{Hibiki Nakada}  %
\email[Email address: ]{nakada27@rcnp.osaka-u.ac.jp}
\affiliation{Research Center for Nuclear Physics (RCNP), Osaka University, Ibaraki, Osaka 567-0047, Japan}

\author{Kazuki Yoshida}  %
\affiliation{Advanced Science Research Center, Japan Atomic Energy Agency, Tokai, Ibaraki 319-1195, Japan}
\author{Kazuyuki Ogata}  %
\affiliation{Department of Physics, Kyushu University, Fukuoka 819-0395, Japan}
\affiliation{Research Center for Nuclear Physics (RCNP), Osaka University, Ibaraki, Osaka 567-0047, Japan}

             

\begin{abstract}
\noindent
\textbf{Background}:~The description of deuteron-induced inclusive reactions has been an
important subject in direct nuclear reaction studies and nuclear data science. For proton-induced inclusive processes, the semiclassical distorted wave model (SCDW) is one of the most successful models based on quantum mechanics.

\noindent
\textbf{Purpose}:~We improve SCDW for deuteron-induced inclusive processes and clarify the importance of the proper treatment of the kinematics of the deuteron inside a nucleus.

\noindent
\textbf{Methods}:~The double differential cross section (DDX) of the inclusive deuteron-emission process $(d,d^{\prime}x)$ is described by one-step SCDW.
The changes in the kinematics due to the distortion effect, the refraction effect, is taken into account by the local semiclassical approximation (LSCA).

\noindent
\textbf{Results}:~The calculated DDXs of $(d,d^{\prime}x)$ reasonably reproduce experimental data in the small energy-transfer region and at forward and middle angles with some exceptions. The angular distributions of $(d,d^{\prime}x)$ are improved by including the refraction effect. 

\noindent
\textbf{Conclusion}:~The proper treatment of the changes in the kinematics of the deuteron inside a nucleus is necessary in describing the ($d$,$d'x$) reaction. The effect of the changes on the DDX of $(d,d^{\prime}x)$ is significant compared to on the proton-induced inclusive process $(p,p^{\prime}x)$ because of the stronger distortion effect on the deuteron.

\end{abstract}


\maketitle


\section{INTRODUCTION}\label{Sec_1}

Deuteron has the smallest binding energy among all stable nuclei. As originated from the idea of Butler~\cite{butler1950}, the weakly-bound nature of deuteron has been utilized for carrying out one-nucleon transfer reactions to study the single-particle (s.p.) structure of nuclei~\cite{timofeyuk2020}. Furthermore, deuteron-induced reactions have opened many physics cases to reveal three-body dynamics of reaction systems in which a fragile nucleus is involved~\cite{Kamimura1986,austern1987d,deltuva2009,upadhyay2012,ogata2016,potel2017}. Roles of deuteron breakup channels, in which proton and neutron are in continuum states, have intensively been investigated.

The fragileness of deuteron is also important for nuclear data science. The international fusion materials irradiation facility (IFMIF)~\cite{Moeslang2006}, which aims at using the inclusive $(d,nx)$ reaction at 40~MeV as an intense neutron source, is one of the most well-known international scientific projects using deuteron accelerator. The central idea of IFMIF is that the incident deuteron is broken up by interacting with the target and intense neutron with about half the deuteron incident energy is emitted; statistical decay after forming a compound nucleus is also considered to contribute to the neutron emission for large energy transfer. Quite recently, an integrated code system describing deuteron-induced reactions, which is designated as DEURACS, has been constructed and successfully applied to analysis of $(d,nx)$ reaction data~\cite{nakayama2016,nakayama2020,nakayama2021}. It was found that the description of deuteron breakup channels is of crucial importance for accurately evaluating the amount of the emitted neutron, its angular and energy distribution in particular.

From the viewpoint of direct nuclear reaction study, the most challenging part for describing ($d,nx$) is the deuteron breakup with exciting the target nucleus A, which is called the nonelastic breakup (NEB). NEB contains a huge number of final states of A and it is almost impossible to describe each nuclear state accurately. DEURACS employs the Glauber model~\cite{glauber1959} to circumvent the difficulty; the eikonal and adiabatic approximations allow one to describe NEB as a combination of neutron elastic and proton nonelastic processes, and the latter can easily be evaluated by using the closure property of the proton scattering matrix~\cite{Hussein1985,hencken1996a}. The validity of the Glauber model is, however, rather questionable at low incident energy and/or for large momentum and energy transfer. In fact, the agreement between the result of DEURACS and experimental data for ($d,nx$) at middle emission angles is slightly flawed compared with that at forward angles~\cite{nakayama2021}. Although the neutron emission cross section is forward-peaked and the {\lq\lq}deviation'' is not very serious for practical use, the description of NEB of deuteron without using the eikonal and adiabatic approximations will be an important subject of nuclear reaction study. Recently, the Ichimura-Austern-Vincent (IAV) model~\cite{ichimura1985} has successfully been applied to NEB in several cases~\cite{potel2017,lei2015,lei2015a}. It should be noted, however, that in the IAV model for ($d,nx$), the kinematics of the neutron are not affected at all by the nonelastic processes for which the proton and A undergo. In this sense, the three-body kinematics are not treated in a fully consistent manner in the IAV model.

On the other hand, for proton-induced inclusive processes, ($p,p'x$), several quantum-mechanical models~\cite{Feshbach1980,nishioka1988,tamura1982,Luo1991} have been developed and successfully reproduced experimental data. Among them, the semiclassical distorted wave model (SCDW)~\cite{Luo1991,kawai1992,watanabe1999,ogata1999,weili1999,ogata2002} has no free adjustable parameter and allows a simple intuitive picture of ($p,p'x$). The original SCDW adopted the local Fermi-gas model (LFG) for initial and final nuclear s.p. states. Although LFG will be totally unrealistic for modeling specific nuclear states, it will reasonably describe the total response of a nucleus to which many initial and final states contribute. It should be noted that, in SCDW, there is no kinematical assumption or restriction for the reaction particles. This idea for treating processes via a huge number of nuclear states is expected to work also for deuteron-induced reactions. Note that the latest version of SCDW adopts the Wigner transform of one-body density matrices calculated with a s.p. model for nuclei~\cite{weili1999} instead of LFG; for reducing numerical task, we use LFG in this work.

The main purpose of this study is to extend SCDW to deuteron-induced inclusive processes. Although our ultimate goal is to describe ($d,nx$), as the first step, we focus on the inclusive deuteron-emission process ($d,d'x$). We assume for simplicity that scattering waves of the incoming and outgoing deuteron can be described with a phenomenological optical potential, meaning that deuteron breakup channels are not explicitly treated but net loss of probability is implicitly taken into account by the absorption effect of the deuteron optical potential. On the other hand, as in SCDW studies on ($p,p'x$), we respect kinematics of deuteron inside a nucleus, by using the local semiclassical approximation (LSCA) \cite{Luo1991} to the deuteron distorted waves. We clarify how the proper treatment of the {\lq\lq}refraction'' of deuteron by the distorting potential is important to describe ($d,d'x$) experimental data. We include only the one-step process and mainly discuss the small energy-transfer region.\par
The construction of this paper is as follows. In Sec.~\ref{Sec_2} we describe SCDW for the inclusive $(d,d^{\prime}x)$ reaction, applying LFG and LSCA. In Sec.~\ref{Sec_3} we compare the calculated DDXs of the inclusive $(d,d^{\prime}x)$ reaction with experimental data and demonstrate the effect of nuclear refraction. Finally, a summary is given in Sec.~\ref{Sec_4}.

\section{FORMALISM}\label{Sec_2}
We describe the inclusive $(d,d^{\prime}x)$ reaction by one-step SCDW. 
The foundation of SCDW is the DWBA series expansion of the transition matrix (\textit{T} matrix). The \textit{T}-matrix element, for which the target nucleus is excited from the initial s.p. state $\phi_{\alpha}$ to the final one $\phi_{\beta}$, is given by
\begin{align}
T_{\beta \alpha}={\Braket{\chi^{(-)}_{f}(\bm{r}_{0})\phi_{\beta}(\bm{r})|v(\bm{r}_{0}-\bm{r})|\chi^{(+)}_{i}(\bm{r}_0)\phi_{\alpha}(\bm{r})}},
\label{eq_T-matrix}
\end{align}
where $\bm{r}_{0}$ and $\bm{r}$ are the coordinates of the incident deuteron and the nucleon inside the target, respectively.
$\chi_{i}$ ($\chi_{f}$) is the distorted wave for the deuteron in the initial (final) state. The superscripts $(+)$ and $(-)$ denote the outgoing and incoming boundary conditions for $\chi$, respectively. $v$ is the effective interaction between the deuteron and the target nucleon. The double differential cross section (DDX) for the emitted deuteron energy $E_f$ and the solid angle $\Omega_f$ is given by

\begin{align}
\frac{\partial^2\sigma}{\partial E_{f}\partial\Omega_{f}}= C\frac{k_{f}}{k_{i}}\sum_{\alpha, \beta}|T_{\beta \alpha}|^2\delta(E_{i}+\varepsilon_{\alpha}-E_{f}-\varepsilon_{\beta}), 
\label{eq_DDX_DWBA}
\end{align}
where $C=4\mu^2/(2\pi\hbar^2)^2$, $E_i$ is the deuteron incident energy, $\mu$ is the reduced mass between the deuteron and the target nucleus and $k_{i}$ ($k_{f}$) is the asymptotic momentum of the incident (emitted) deuteron. $\varepsilon_{\gamma}~(\gamma=\alpha~\mathrm{or}~\beta)$ is the kinetic energy of the target nucleon. The summation is taken over all the initial and the final s.p. states, $\alpha$ and $\beta$, which are relevant to the inclusive $(d,d^{\prime}x)$ reaction. 
On expanding the squared modulus in Eq.~(\ref{eq_DDX_DWBA}), one obtains 
\begin{align}
\frac{\partial^2\sigma}{\partial E_{f}\partial\Omega_{f}}=C\frac{k_{f}}{k_{i}}\int d\bm{r}_{0}d\bm{r}\chi^{\ast(-)}_{f}(\bm{r}_{0})v(\bm{r}_{0}-\bm{r})\chi^{(+)}_{i}(\bm{r}_{0}) \nonumber \\
\times\int d\bm{r}_0^{\prime}d\bm{r}^{\prime}\chi^{(-)}_{f}(\bm{r}^{\prime}_{0})v^{\ast}(\bm{r}^{\prime}_{0}-\bm{r}^{\prime})\chi^{\ast(+)}_{i}(\bm{r}^{\prime}_{0})K(\bm{r},\bm{r}^{\prime}),  
\label{eq_DDX_r0r}
\end{align}
where the kernel $K(\bm{r},\bm{r}^{\prime})$ is define by
\begin{align}
  K(\bm{r},\bm{r}^{\prime}) 
  &\equiv \sum_{\alpha}\phi_{\alpha}(\bm{r})\phi^{\ast}_{\alpha}(\bm{r}^{\prime})\sum_{\beta}\phi^{\ast}_{\beta}(\bm{r})\phi_{\beta}(\bm{r}^{\prime}) \nonumber \\
&\times\delta(E_{i}+\varepsilon_{\alpha}-E_{f}-\varepsilon_{\beta}).   
\label{eq_K_rrp}
\end{align}
When a large number of s.p. states are involved, $K(\bm{r},\bm{r}^{\prime})$ becomes a short-ranged function of $|\bm{r}-\bm{r}^{\prime}|$~\cite{watanabe1999,kawai1992,watanabe1993}. The center-of-mass and relative coordinates of the $d$-$N$ system, $\bm{R}$ and $\bm{s}$, respectively,
are given by
\begin{align}
  \bm{R}&=\frac{A_{d}}{A_{d}+1}\bm{r}_0+\frac{1}{A_{d}+1}\bm{r},
  \label{eq_R_r0r} \\
  \bm{s}&=\bm{r}_{0}-\bm{r}.
  \label{eq_s_r0r}
\end{align}
Inversely, $\bm{r}_{0}$ and $\bm{r}$ are written as
\begin{align}
    \bm{r}_{0}=\bm{R}+\frac{1}{A_{d}+1}\bm{s},
    \label{eq_r0_Rs} \\
    \bm{r}=\bm{R}-\frac{A_d}{A_{d}+1}\bm{s},
    \label{eq_r_Rs}
\end{align}
where $A_d$ is the mass number of deuteron, i.e.,  $A_d=2$. With the coordinates $\bm{R}$ and $\bm{s}$, one can rewrite Eq.~(\ref{eq_DDX_r0r}) as
\begin{align}
\frac{\partial^2\sigma}{\partial E_{f}\partial\Omega_{f}}
&=C\frac{k_{f}}{k_{i}}
\int d\bm{R}\,d\bm{s}\,d\bm{R}^{\prime}\,d\bm{s}^{\prime} \nonumber \\
&\times
\chi^{\ast(-)}_{f}(\bm{R}+\bm{s}/3)v(\bm{s})\chi^{(+)}_{i}(\bm{R}+\bm{s}/3) \nonumber \\
&\times
\chi^{(-)}_{f}(\bm{R}^{\prime}+\bm{s}^{\prime}/3)v^{\ast}(\bm{s}^{\prime})\chi^{\ast(+)}_{i}(\bm{R}^{\prime}+\bm{s}^{\prime}/3) \nonumber \\ 
&\times K(\bm{R},\bm{s},\bm{R}^{\prime},\bm{s}^{\prime}),  
\label{eq_DDX_Rs}
\end{align}
where
\begin{align}
  K(\bm{R},\bm{s},\bm{R}^{\prime},\bm{s}^{\prime})
  &=\sum_{\alpha}\phi_{\alpha}(\bm{R}-2\bm{s}/3)\phi^{\ast}_{\alpha}(\bm{R}^{\prime}-2\bm{s}^{\prime}/3) \nonumber \\
  &\times\sum_{\beta}\phi^{\ast}_{\beta}(\bm{R}-2\bm{s}/3)\phi_{\beta}(\bm{R}^{\prime}-2\bm{s}^{\prime}/3) \nonumber \\
  &\times\delta(E_{i}+\varepsilon_{\alpha}-E_{f}-\varepsilon_{\beta}).  
\label{eq_K_Rs}
\end{align}

Here, we make two approximations to Eq.~(\ref{eq_DDX_Rs}). One is LFG for nuclear states and the other is LSCA for the distorted waves as mentioned in Sec.~\ref{Sec_1}. In LFG, $\phi_{\gamma}~(\gamma=\alpha~\mathrm{or}~\beta)$ is approximated by the plane wave with momentum $k_{\gamma}$ within a smaller cell-size of $|\bm{s}|$ than the range of $v$. The summation of $\gamma$ is then expressed as an integral over $k_{\gamma}$, where the threshold momentum between channels $\alpha$ and $\beta$ is the local Fermi momentum $k_{F}(\bm{R})$, which is related to
the nuclear density $\rho(\bm{R})$ through
\begin{align}
   \rho(\bm{R})=4\frac{4\pi}{3}\frac{k^{3}_F(\bm{R})}{(2\pi)^3}.
  \label{eq_rho_kF}
\end{align}
In LSCA, the short-range propagation of the distorted wave $\chi_{c}~(c=i~\mathrm{or}~f)$ from a reference point $\bm{R}$ is approximated by the plane wave, i.e.,
\begin{align}
    \chi_{c}(\bm{R}+\bm{s}/3)\simeq\chi_{c}(\bm{R})e^{i\bm{k}_{c}(\bm{R})\cdot\bm{s}/3}.
    \label{eq_LSCA_chi}
\end{align}
This approximation is valid because the range of the $d$-$N$ interaction $v$ is short and therefore only a small $\bm{s}$ is relevant to the reaction.  
In Eq.~(\ref{eq_LSCA_chi}), $\bm{k}_{c}(\bm{R})$ is the real local momentum of the deuteron. 
The direction of $\bm{k}_{c}(\bm{R})$ is taken to be the same as that of the flux of the distorted wave $\chi_c(\bm{R})$. The magnitude of $\bm{k}_{c}(\bm{R})$ is given by that of the real part of the complex local momentum $\bm{K}_{c}(\bm{R})$ satisfying the local energy conservation~\cite{watanabe1999}
\begin{align}
\frac{\hbar^{2}k_{c}^2}{2\mu}=\frac{\hbar^{2}K^{2}_{c}(\bm{R})}{2\mu}+ U_{c}(\bm{R}),   
\label{eq_local_conv}
\end{align}
where $U_c(\bm{R})~(c=i~\mathrm{or}~f)$ is a complex distorting potential for the deuteron. In present calculation, we use the optical potential for $U_{c}(\bm{R})$ as mentioned in Sec.~\ref{sub_A}.\par
 Using LFG and LSCA, one can rewrite Eq.~(\ref{eq_DDX_Rs}) as
\begin{align}
\frac{\partial^2\sigma}{\partial E_{f}\partial\Omega_{f}}
&= \frac{C}{(2\pi)^{3}}\frac{k_{f}}{k_{i}}\int d\bm{R}|\chi^{(-)}_{f}(\bm{R})|^{2}|\chi^{(+)}_{i}(\bm{R})|^{2} \nonumber \\
&\times \int_{k_{\alpha}\leq k_{F}(\bm{R})}d\bm{k}_{\alpha}\int_{k_{\beta}>k_{F}(\bm{R})}d\bm{k}_{\beta} \nonumber \\
&\times \left|\int d\bm{s}v(\bm{s})e^{-i\bm{q}(\bm{R})\cdot\bm{s}}\right|^{2} \nonumber \\
&\times \delta(\bm{k}_{i}(\bm{R})+\bm{k}_{\alpha}-\bm{k}_{f}(\bm{R})-\bm{k}_{\beta}) \nonumber \\
&\times \delta(E_{i}+\varepsilon_{\alpha}-E_{f}-\varepsilon_{\beta}),
\label{eq_DDX_SCDW}
\end{align}
where $\bm{q}(\bm{R})$ is the local momentum transfer defined by $\bm{k}_{i}(\bm{R})-\bm{k}_{f}(\bm{R})$. 
In Eq.~(\ref{eq_DDX_SCDW}), Dirac's delta functions and the ranges of the integrations, $k_{\alpha}\leq k_{F}(\bm{R})$ and $k_{\beta}>k_{F}(\bm{R})$, guarantee that the $d$-$N$ elementary process satisfies the Pauli principle and the energy and local momentum conservation in the $(d,d^{\prime}x)$ reaction.
We make the on-the-energy-shell approximation to the squared modulus of the matrix element of $v$:
\begin{align}
\frac{\mu^2_{dN}}{(2\pi\hbar^2)^2}\left|\int d\bm{s}v(\bm{s})e^{-i\bm{q}(\bm{R})\cdot\bm{s}}\right|^{2}\simeq\left(\frac{d\sigma_{dN}}{d\Omega}\right)_{\theta_{dN}(\bm{R}),E_{dN}(\bm{R})},  
\label{eq_t-mat_appro}
\end{align}
where $\mu_{dN}$ is the reduced mass of the $d$-$N$ system. $\theta_{dN}(\bm{R})$ is the local $d$-$N$ scattering angle between the initial relative momentum $\bm{\kappa}(\bm{R})$ and the final one $\bm{\kappa}^{\prime}(\bm{R})$, which are defined by
\begin{align}
    \bm{\kappa}(\bm{R})
    &\equiv\frac{1}{A_d+1}\bm{k}_{i}(\bm{R})-\frac{A_d}{A_d+1}\bm{k}_{\alpha},
    \label{eq_Kappa_kika} \\
    \bm{\kappa}^{\prime}(\bm{R})
    &\equiv\frac{A_{d}}{A_d+1}\bm{k}_{f}(\bm{R})-\frac{1}{A_d+1}\bm{k}_{\beta}.
    \label{eq_kappap_kfkb}
\end{align}
The local $d$-$N$ scattering energy $E_{\bm{R}}$ is defined by
\begin{align}
    E_{dN}(\bm{R})=\frac{\hbar^2\kappa^2(\bm{R})}{2\mu_{dN}}.
   \label{eq_EdN}
\end{align}
By substituting Eq.~(\ref{eq_t-mat_appro}) for Eq.~(\ref{eq_DDX_SCDW}) and integrating over $\bm{k}_{\beta}$, one obtains the following closed form of the DDX of the inclusive $(d,d^{\prime}x)$ reaction: 
\begin{align}
\frac{\partial^2\sigma}{\partial E_{f}\partial\Omega_{f}}
&= \left[\frac{A_{d}A}{A_{d}+A}\right]^2\frac{k_{f}}{k_{i}}\int d\bm{R} \nonumber \\
&\times|\chi^{(-)}_{f}(\bm{R})|^{2}|\chi^{(+)}_{i}(\bm{R})|^{2}\left[\frac{\partial^2\sigma}{\partial E_{f}\partial\Omega_{f}}\right]_{\bm{R}}\rho(\bm{R}),
\label{eq_DDX_ddx}
\end{align}
where $A$ is the mass number of the target nucleus. 
The DDX of the elementary process averaged over $\bm{k}_\alpha$ at $\bm{R}$ in the Fermi sphere characterized by $k_F^3(\bm{R})$ is given by
\begin{align}
\left[\frac{\partial^2\sigma}{\partial E_{f}\partial\Omega_{f}}\right]_{\bm{R}}
&= \frac{1}{(4\pi/3)k^3_{F}(\bm{R})}\left[\frac{A_{d}+1}{A_{d}}\right]^2 \nonumber \\
&\times \int_{k_{\alpha}\leq k_{F}(\bm{R})}d\bm{k}_{\alpha}\left(\frac{d\sigma_{dN}}{d\Omega}\right)_{\theta_{dN}(\bm{R}),E_{dN}(\bm{R})} \nonumber \\
&\times \delta(E_{i}+\varepsilon_{\alpha}-E_{f}-\varepsilon_{\beta}).
\label{eq_DDX_ave}
\end{align}

LSCA can incorporate the distortion effect on the kinematics of the incident and emitted particles. This effect can be regarded as refraction due to the distorting potential because the direction of the local momentum changes continuously as a function of ${\bm R}$. To clarify the refraction effect, we also consider the asymptotic momentum approximation~(AMA), which replaces $\bm{k}_c(\bm{R})$ with $\bm{k}_c$, i.e., $\bm{k}_c(\bm{R})\rightarrow \bm{k}_c$ in Eq.~(\ref{eq_LSCA_chi}). 
\begin{align}
    \chi_{c}(\bm{R}+\bm{s}/3)\simeq\chi_{c}(\bm{R})e^{i\bm{k}_{c}\cdot\bm{s}/3}.
    \label{eq_AMA_chi}
\end{align}
If AMA shown in Eq.~(\ref{eq_AMA_chi}) is used instead of LSCA, $\bm{k}_c(\bm{R})$ is constant with respect to $\bm{R}$ in Eqs.~(\ref{eq_local_conv})-(\ref{eq_DDX_ave}). The effect of the refraction is discussed in Sec.~\ref{sub_C}. The validity of LSCA and AMA is given in Appendix.\par

\begin{figure*}[htbp]
    \centering
    \includegraphics[width=0.9\hsize]{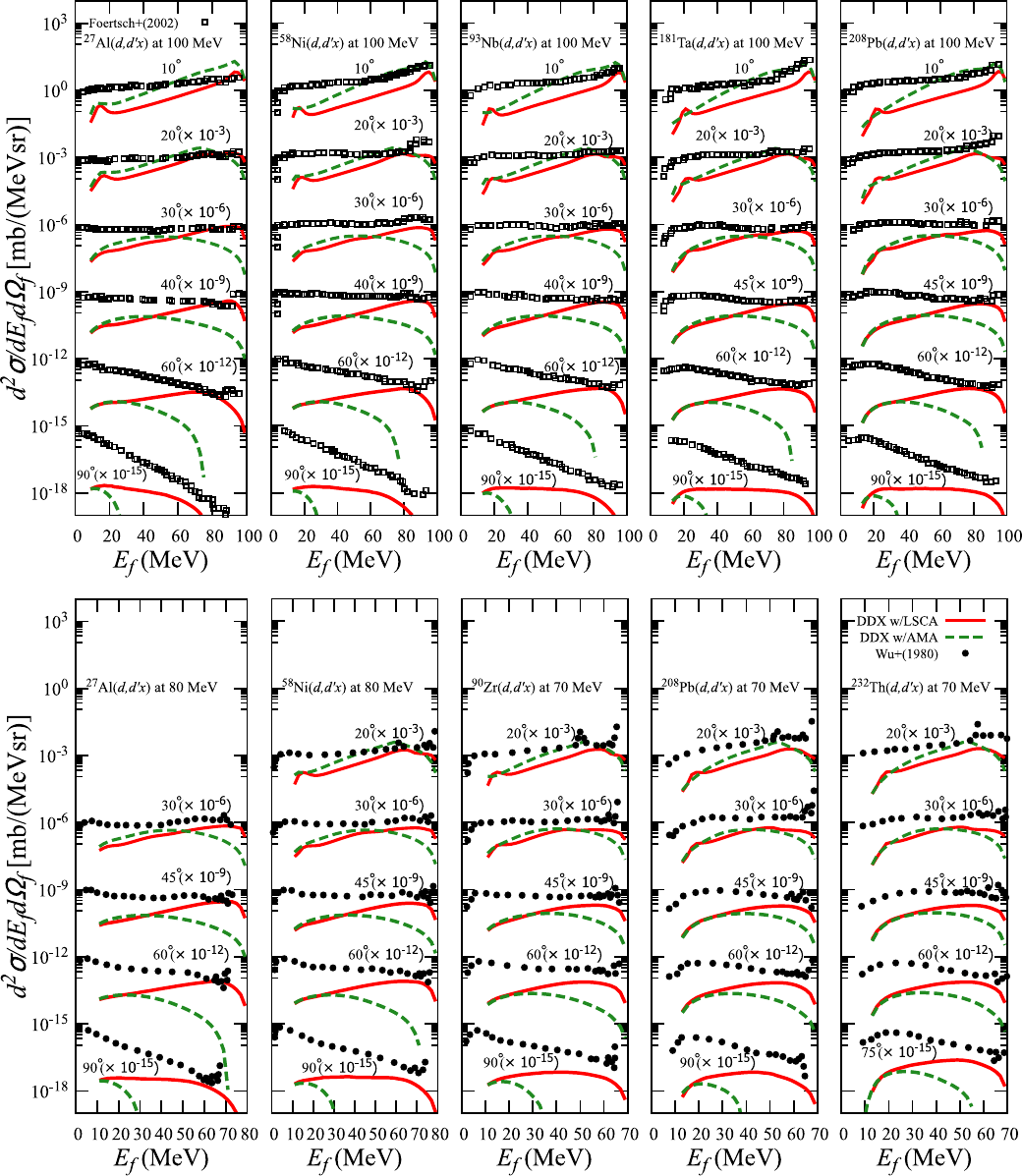}
    \caption{Comparison of the experimental data and calculated DDXs of the inclusive $(d, d^{\prime}x)$ reaction on $^{27}\mathrm{Al}$ and $^{58}\mathrm{Ni}$ at 100 MeV and 80 MeV, $^{208}\mathrm{Pb}$ at 100 MeV and 70 MeV, $^{93}\mathrm{Nb}$ and $^{181}\mathrm{Ta}$ at 100 MeV, and $^{90}\mathrm{Zr}$ and $^{232}\mathrm{Th}$ at 70 MeV for different deuteron emission angles. The solid (dashed) lines represent the DDXs with LSCA (AMA). The experimental data at 100 MeV are taken from Ref. \cite{Fortsch2002} and those at 80 and 70 MeV are from Ref. \cite{Wu1979}.}
    \label{fig:DDXs_E}
\end{figure*}
\section{RESULTS AND DISCUSSION}\label{Sec_3}
    

\subsection{Numerical inputs}\label{sub_A}
We assume the Woods-Saxon shaped global optical potential by An and Cai~\cite{An2006} for the distorting potential $U_{c}(\bm{R})$ of the deuteron scattering off target nuclei.
The effect of the nonlocality of the deuteron distorting potentials is taken into account by multiplying the scattering waves by the Perey factor \cite{Perey1962} $F_{c}(R)=[1-\mu\beta^2/(2\hbar^2)U_c(R)]^{-1/2}$, where $\mu$ is the reduced mass of the deuteron and the target. The range of nonlocality $\beta$ for the deuteron is taken to be 0.54 \cite{TWOFNR}.
We assume the Woods-Saxon form for the nuclear density as
\begin{align}
\rho(R) = \frac{\rho_0}{1+\mathrm{exp}\left(\frac{R-R_\rho}{a_\rho}\right)},
\label{eq_rho_WS}
\end{align}
where the radial parameter is given by $R_\rho=r_\rho A^{1/3}$ with $r_\rho=1.15$~fm, the diffuseness parameter is set to $a_\rho=0.5$~fm as in Ref.~\cite{Luo1991}, and $A$ being the mass number of the target nucleus.
The constant $\rho_0$ is determined to normalize the integrated value of $\rho (R)$ to $A$. The local Fermi momentum is calculated from the nucleon density as in Eq.~(\ref{eq_rho_kF}).\par
For the free $d$-$N$ scattering cross section, we use the numerical table fitted with several Gaussian functions to reproduce the experimental data of $p$-$d$ scattering from 5 to 800 MeV~\cite{chazono2022a}. In this table, the cross section does not diverge at $0^\circ$ because we neglect the Coulomb elastic scattering. For the free $p$-$N$ scattering cross section used in the calculation of the $(p,p^{\prime}x)$ process, we use the nucleon-nucleon \textit{t} matrix by Franey and Love~\cite{Love1981,Franey1985}.


\subsection{Results of the SCDW calculation for $(d,d^{\prime}x)$  reactions and comparison with data}\label{sub_B}
We show the DDXs of the inclusive $(d,d^{\prime}x)$ reaction calculated with SCDW using LSCA and AMA, and compare them with experimental data. Below we discuss the DDXs as a function of the emission energy $E_f$ of the deuteron with fixed $\Omega_f$. 
Figure~\ref{fig:DDXs_E} shows the calculated DDXs as a function of $E_f$ at the several laboratory scattering angles $\theta$. The solid (dashed) lines represent the DDXs using LSCA (AMA). The DDXs are calculated for $^{27}\mathrm{Al}$ and $^{58}\mathrm{Ni}$ at the incident energy $E_i = 100$ and $80$~MeV, $^{208}\mathrm{Pb}$ at 100 and 70 MeV, and $^{93}\mathrm{Nb}$ and $^{181}\mathrm{Ta}$~($^{90}\mathrm{Zr}$ and $^{232}\mathrm{Th}$) at 100 MeV (70 MeV). The experimental data at 100 MeV are taken from Ref.~\cite{Fortsch2002} and those at $E_i = 80$ and $70$ MeV are taken from Ref.~\cite{Wu1979}. In the experimental data at 80 and 70 MeV, the sharp increase at very large $E_f$ is due to the elastic scattering events.\par
The contribution of multi-step processes becomes larger as the energy transfer $\omega$ or $\theta$ increases, as discussed in $(p, nx)$ reaction in Ref.~\cite{ogata2002}. For this reason, we focus only on the regions: $\omega = E_i-E_f \lesssim 15$~MeV and $\theta \lesssim 60^{\circ}$ in Fig.~\ref{fig:DDXs_E}, where one-step process is expected to be dominant. The calculated DDXs with LSCA reproduce the experimental data reasonably well, although in some cases undershooting is found. It turns out that the undershooting is pronounced at smaller $E_i$ or for heavier target nuclei and is particularly noticeable in the DDXs on $^{58}\mathrm{Ni}$ at 100 MeV. Further investigation is needed on this underestimation issue.\par
In the small energy-transfer region, $\omega \lesssim 15$~MeV, the DDXs with AMA decrease rapidly as $\theta$ increases and show too strong angular dependence compared to the data. 
In contrast to the AMA results, the DDXs with LSCA show better agreements with the data in the angular dependence.
At $\theta\geq 30^{\circ}$, DDXs with LSCA are much closer to the experimental data than those with AMA as discussed in Sec.~\ref{sub_C}. These results imply that the inclusion of the nuclear refraction, i.e., the change in the kinematics of the deuteron inside the nucleus, is necessary to describe the angular dependence of the experimental data.


\subsection{The effect of the refraction}\label{sub_C}
\begin{figure}[htbp]
    \centering
    \includegraphics[width=0.9\hsize]{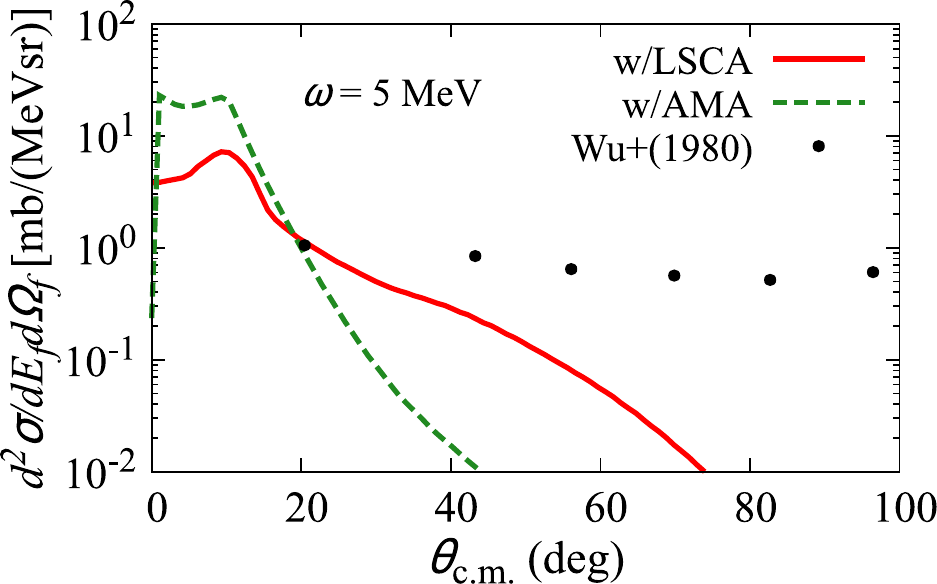}
    \caption{DDXs of the $^{58}\mathrm{Ni}(d,d^{\prime}x)$ at 80 MeV as a function of the scattering angle with $\omega=5$ MeV. The solid (dashed) line corresponds to the calculation with LSCA (AMA). The experimental data are taken from Ref.~\cite{Wu1979}.}
    \label{fig:DDX_omega5}
\end{figure}

Below we discuss the angular dependence of the nuclear refraction on the DDX  of the $^{58}\mathrm{Ni}(d,d^{\prime}x)$ at 80 MeV. Figure~\ref{fig:DDX_omega5} shows the calculated DDXs as a function of the center-of-mass scattering angle $\theta_{\mathrm{c.m.}}$ with $\omega=5$ MeV and the experimental data in Ref.~\cite{Wu1979}. The solid (dashed) line represents the DDX with LSCA (AMA).
It is shown that the DDX with LSCA gives moderate angular dependence and the results at the middle and backward angles are improved, compared with those with AMA. However, the gap between the result with LSCA and the experimental data remains and the multi-step processes will be necessary to fill it.\par

By comparing the calculated DDXs using LSCA and AMA, one can see that there are mainly two effects of the nuclear refraction. One is the extension of the allowed region of $\theta_{\mathrm{c.m.}}$. For the DDX with AMA, the ($d, d^{\prime}x$) reaction is only allowed to about $42^{\circ}$. On the other hand, the DDX with LSCA extends up to $\theta_{\mathrm{c.m.}}=73^{\circ}$ and does not drop off at very small $\theta_{\mathrm{c.m.}}$. 
This is because kinematics forbidden in AMA become allowed in LSCA by the refraction of the momentum of the deuteron in the target nucleus. It should be noted that, as one may find from Eq.~(\ref{eq_DDX_ddx}), the ${\bm R}$ dependence of the kinematics of the deuteron due to the refraction dictates the  averaged local cross section of the $d$-$N$ elementary process. In other words, whether the $d$-$N$ process can take place or not depend on ${\bm R}$ through $\bm{q}({\bm R})$ and $k_F({\bm R})$.\par
\begin{figure}[htbp]
 \centering
 \includegraphics[width=1.0\hsize]{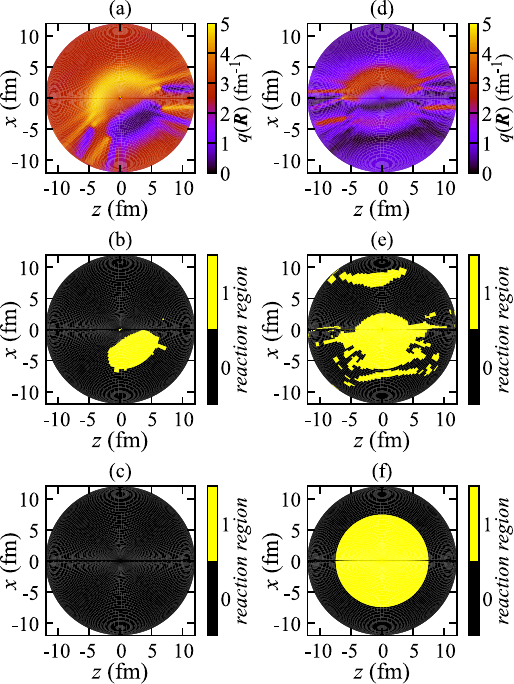}   
 \caption{(a) The local momentum transfer with LSCA of $^{58}\mathrm{Ni}(d,d^{\prime}x)$ at 80 MeV with $\omega = 5$ MeV at $\theta_{\mathrm{c.m.}} = 60^{\circ}$. (b) The reaction region with LSCA. The color bar with the value of ``1'' means the region in which the $d$-$N$ process is allowed, while with the value of ``0'' means the region in which the process is not allowed. (c) Same as (b) but with AMA. (d), (e) and (f) same as (a), (b) and (c), respectively, but at $\theta_{\mathrm{c.m.}} =15^{\circ}$.}
\label{fig:qR}
\end{figure}

To see this more clearly, we show in Fig.~\ref{fig:qR}(a) $\bm{q}(\bm{R})$ calculated with LSCA, and in Figs.~\ref{fig:qR}(b) and (c) the kinematically-allowed reaction regions of the elementary process with LSCA and AMA, respectively, at $\theta_{\mathrm{c.m.}}=60^{\circ}$ of the Fig.~\ref{fig:DDX_omega5} case. 
In Figs.~\ref{fig:qR}(b) and (c), a color bar with the value of ``1'' indicates the regions where the $d$-$N$ processes are allowed while ``0'' indicates the regions where the $d$-$N$ processes are not allowed. In AMA, which does not include the nuclear refraction, $\bm{q}(\bm{R})$ is the same as the asymptotic momentum transfer $\bm{q}$. Figure~\ref{fig:qR}(c) shows that there are no kinematically-allowed reaction regions; it is found that this is because $q$ is too large to allow the elementary process. On the other hand, with LSCA, there is a region in which the $d$-$N$ process is allowed because $\bm{q}(\bm{R})$ is dispersed by the nuclear refraction and may have smaller values.\par
The other effect of refraction is the decrease in the DDX at forward angles. This is because LSCA makes kinematically-allowed reaction regions narrower. 
Figures~\ref{fig:qR}(d), (e), and (f) are the same as Figs.~\ref{fig:qR}(a), (b), and (c), respectively, but at $\theta_{\mathrm{c.m.}}=15^{\circ}$. In Fig.~\ref{fig:qR}(f), one can see that the $d$-$N$ process is kinematically allowed in the very broad region when AMA is used. On the other hand, in Fig.~\ref{fig:qR}(e), the reaction region with LSCA becomes narrower than that with AMA. This is because $\bm{q}(\bm{R})$ is dispersed and can have too large values, as in Fig.~\ref{fig:qR}(d), to kinematically allow the $d$-$N$ process.\par
\begin{figure}[t]
    \includegraphics[width=0.9\hsize]{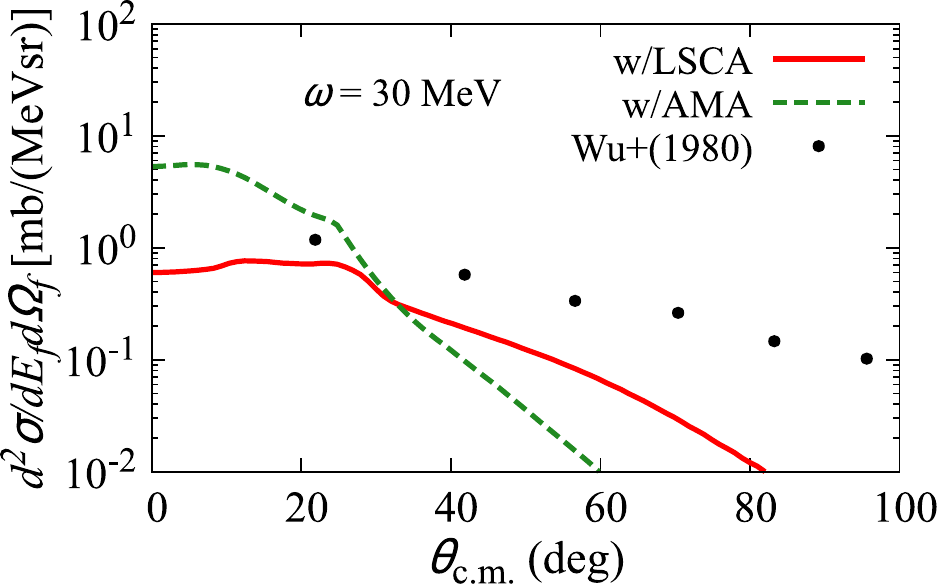}
    \caption{Same as Fig.~\ref{fig:DDX_omega5} but for $\omega=30$ MeV.}
    \label{fig:DDX_omega30}
\end{figure}

From these results, we conclude that the two effects of the nuclear refraction can be understood as the changes in the kinematically-allowed reaction regions associated with the dispersion of $\bm{q}(\bm{R})$.
Figure~\ref{fig:DDX_omega30} is the same as Fig.~\ref{fig:DDX_omega5} but with $\omega = 30$~MeV. In Fig.~~\ref{fig:DDX_omega30}, we can see that the two effects of refraction remain even when $\omega$ is large. The angular distribution of the DDX with LSCA gives a better agreement with the experimental data than that with AMA, as in the case of $\omega=5$~MeV shown in Fig.~\ref{fig:DDX_omega5}.\par
\begin{figure}[htbp]
    \includegraphics[width=0.9\hsize]{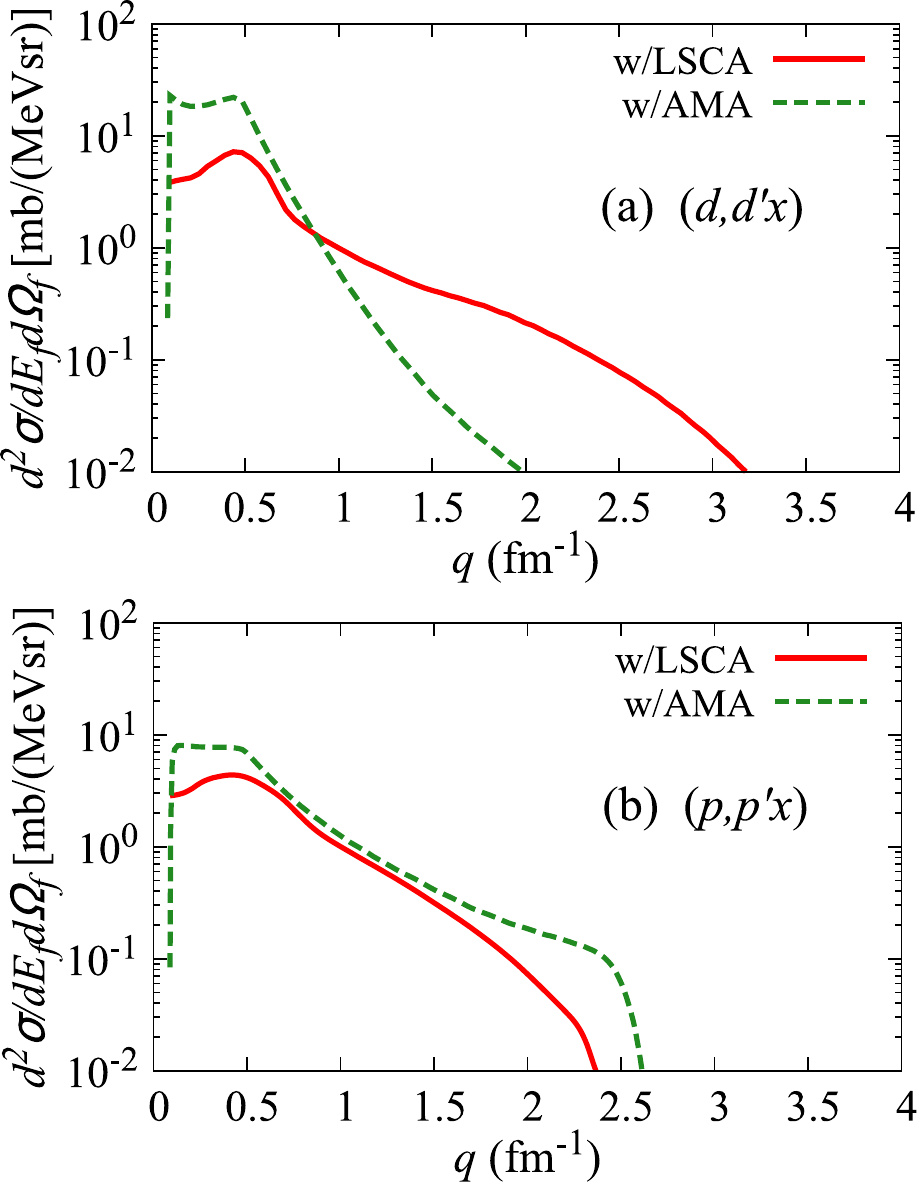}
    \caption{(a)Same as Fig.~\ref{fig:DDX_omega5} but horizontal axis is the momentum transfer. (b) Same as (a) but of $^{58}\mathrm{Ni}$$(p,p^{\prime}x)$.}
   \label{DDXs_q_E50}
\end{figure}

Figures~\ref{DDXs_q_E50}(a) and (b) show the DDXs, with $\omega = 5$~MeV, of $^{58}\mathrm{Ni}(d,d^{\prime}x)$ at 80 MeV (40 MeV per nucleon) and $^{58}\mathrm{Ni}(p,p^{\prime}x)$ at 40 MeV, respectively. 
One sees that the effects of the refraction are more significant in the $(d,d^{\prime}x)$ reaction than in the $(p, p^{\prime}x)$ reaction.
This is mainly because the distorting potential between the deuteron and the target is deeper than that between the proton and the target. Although the importance of the nuclear refraction has been pointed out in the preceding studies of $(p,p^{\prime}x)$ reactions with SCDW~\cite{Luo1991}, its effect is found to be not very significant. For $(d,d^{\prime}x)$, as shown in Fig.~\ref{fig:DDXs_E}, the nuclear refraction completely changes the behavior of the DDX. To analyze the $(d,d^{\prime}x)$ reaction data, inclusion of the nuclear refraction will be necessary.
\section{SUMMARY}\label{Sec_4}
We have improved SCDW to the inclusive $(d,d^{\prime}x)$ reaction. The calculated DDXs of the $(d,d^{\prime}x)$ were compared with the experimental data of various targets at several deuteron emission angles. Except for some cases, the calculated DDXs with LSCA reasonably reproduce the experimental data in the regions where one-step process is expected to be dominant, i.e., for small energy transfer and at forward angles. On the other hand, the DDXs with AMA, which does not include changes in the kinematics of the deuteron due to the distorting potential, has too strong angular dependence and considerably underestimate the experimental data at $\theta\geq30^{\circ}$. These results imply that the nuclear refraction effect on the $d$-$N$ elementary process is necessary to reproduce the experimental data. \par
We have shown two effects of the refraction by comparing the DDXs with LSCA and AMA as a function of the scattering angle. One is the extension of the kinematically-allowed scattering angles to the backward region. The other is the decrease in the DDX at forward angles. Both effects can be understood by the changes in the regions where the $d$-$N$ elementary processes are allowed. It is confirmed that the refraction effect is more significant on the $(d, d^{\prime}x)$ than on the $(p, p^{\prime}x)$ due to the strong distortion effect on deuteron.\par 
For better agreement with experimental data in the large energy-transfer region, it will be necessary to modify the present SCDW model for multi-step processes. Another future work will be to consider the deuteron breakup, which is not explicitly treated in this study, to describe the inclusive $(d,nx)$ reaction that is important in nuclear data science.

\begin{acknowledgments}
The authors thank Y.~Chazono for providing us with the $d$-$N$ scattering cross section. H.N. and K.O. thank Y.~Watanabe for fruitful discussions. This work has been supported in part by Grants-in-Aid of the Japan Society for the Promotion of Science (Grants No. JP20K14475, No. JP21H00125, and No. JP21H04975). The computation was carried out with the computer facilities at the Research Center for Nuclear Physics, Osaka University. 
\end{acknowledgments}

\appendix*
\section{Validity of LSCA and AMA}

The validity of LSCA in nucleon scattering has been examined in Refs.~\cite{watanabe1999,Minomo2010}. In these papers, it is shown that LSCA works well for the propagation up to about 1.5 fm at energies above 50 MeV. The validity of LSCA for the $\alpha$ particle was also verified in Ref.~\cite{yoshida2016}.
 \begin{figure}[htbp]
\includegraphics[width=0.80\hsize]{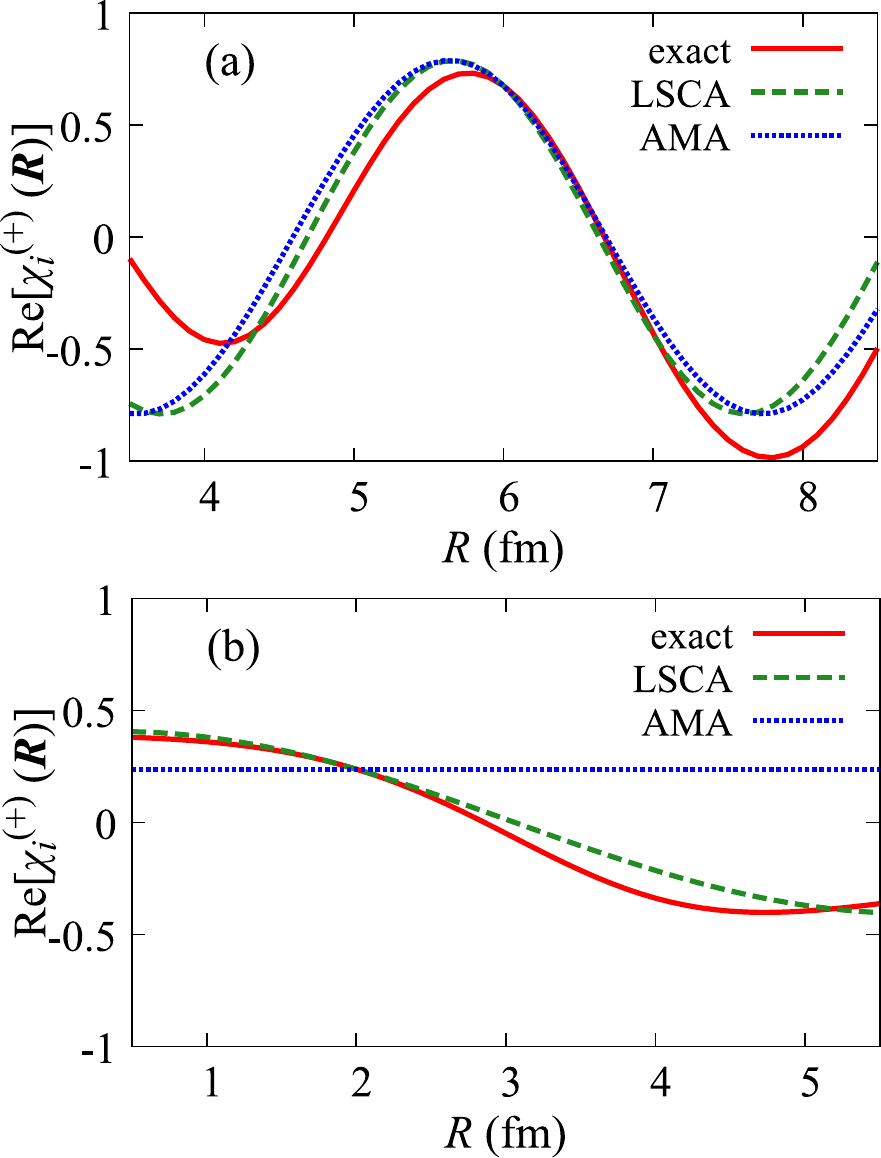}
\caption{The validity of LSCA and AMA.
The real part of the exact $\chi^{(+)}_{i}$ (solid line), with LSCA (dashed line), and with AMA (dotted line) are compared. 
In Figs.~(a) and (b), the propagation in radial direction from $(6~\mathrm{fm},~120^{\circ},~0^{\circ})$ and $(2~\mathrm{fm},~90^{\circ},~0^{\circ})$ in the spherical coordinate representation are shown, respectively.}
\label{fig:A}
\end{figure} 
For the deuteron, however, its validity has not been confirmed. In Fig.~\ref{fig:A}, we examine the validity of LSCA and AMA for the $d$-$^{58}\mathrm{Ni}$ distorted wave $\chi^{(+)}_{i}$ at 50~MeV per nucleon. 
Fig.~\ref{fig:A} shows the propagation in the radial direction from (a) $\bm{R}_{a}\equiv(6~\mathrm{fm}
,~120^{\circ},~0^{\circ})$ and (b) $\bm{R}_{b}\equiv(2~\mathrm{fm}
,~90^{\circ},~0^{\circ})$ in the spherical coordinate representation.
The solid, dashed, and dotted lines show, respectively, the real part of the exact wave function, that with LSCA, and that with AMA.
In Fig.~\ref{fig:A}(a), both approximations reproduce well the propagation up to about 0.7 fm. It should be noted that the range of the interaction between the deuteron and the nucleon is about 2.2 fm, and from the factor $1/(A_{d}+1)=1/3$ for $\bm{s}$ in Eq.~(\ref{eq_r0_Rs}), LSCA and AMA are required to be valid for the propagation up to  about 0.7 fm. In Fig. \ref{fig:A}(b), on the other hand, while LSCA reproduces the propagation of the wave function well, AMA does not. This is because the direction of the propagation direction $\bm{s}$ from $\bm{R}_{b}$ is orthogonal to the asymptotic momentum $\bm{k}_c$ of the deuteron, i.e., $\bm{k}_{c} \cdot \bm{s} = 0$ in Eq.~(\ref{eq_LSCA_chi}). These results show that the kinematics of the deuteron at $\bm{R}_{b}$ are significantly different from the asymptotic ones due to the distorting potential, thus LSCA is essential to trace the deuteron momentum inside the target nucleus.
\nocite{*}

\bibliographystyle{apsrev4-2}
\bibliography{prc1_9}

\providecommand{\noopsort}[1]{}\providecommand{\singleletter}[1]{#1}%
\begin{thebibliography}{38}%
\makeatletter
\providecommand \@ifxundefined [1]{%
 \@ifx{#1\undefined}
}%
\providecommand \@ifnum [1]{%
 \ifnum #1\expandafter \@firstoftwo
 \else \expandafter \@secondoftwo
 \fi
}%
\providecommand \@ifx [1]{%
 \ifx #1\expandafter \@firstoftwo
 \else \expandafter \@secondoftwo
 \fi
}%
\providecommand \natexlab [1]{#1}%
\providecommand \enquote  [1]{``#1''}%
\providecommand \bibnamefont  [1]{#1}%
\providecommand \bibfnamefont [1]{#1}%
\providecommand \citenamefont [1]{#1}%
\providecommand \href@noop [0]{\@secondoftwo}%
\providecommand \href [0]{\begingroup \@sanitize@url \@href}%
\providecommand \@href[1]{\@@startlink{#1}\@@href}%
\providecommand \@@href[1]{\endgroup#1\@@endlink}%
\providecommand \@sanitize@url [0]{\catcode `\\12\catcode `\$12\catcode
  `\&12\catcode `\#12\catcode `\^12\catcode `\_12\catcode `\%12\relax}%
\providecommand \@@startlink[1]{}%
\providecommand \@@endlink[0]{}%
\providecommand \url  [0]{\begingroup\@sanitize@url \@url }%
\providecommand \@url [1]{\endgroup\@href {#1}{\urlprefix }}%
\providecommand \urlprefix  [0]{URL }%
\providecommand \Eprint [0]{\href }%
\providecommand \doibase [0]{https://doi.org/}%
\providecommand \selectlanguage [0]{\@gobble}%
\providecommand \bibinfo  [0]{\@secondoftwo}%
\providecommand \bibfield  [0]{\@secondoftwo}%
\providecommand \translation [1]{[#1]}%
\providecommand \BibitemOpen [0]{}%
\providecommand \bibitemStop [0]{}%
\providecommand \bibitemNoStop [0]{.\EOS\space}%
\providecommand \EOS [0]{\spacefactor3000\relax}%
\providecommand \BibitemShut  [1]{\csname bibitem#1\endcsname}%
\let\auto@bib@innerbib\@empty
\bibitem [{\citenamefont {Butler}(1950)}]{butler1950}%
  \BibitemOpen
  \bibfield  {author} {\bibinfo {author} {\bibfnamefont {S.~T.}\ \bibnamefont
  {Butler}},\ }\href {https://doi.org/10.1103/PhysRev.80.1095.2} {\bibfield
  {journal} {\bibinfo  {journal} {Phys. Rev.}\ }\textbf {\bibinfo {volume}
  {80}},\ \bibinfo {pages} {1095} (\bibinfo {year} {1950})}\BibitemShut
  {NoStop}%
\bibitem [{\citenamefont {Timofeyuk}\ and\ \citenamefont
  {Johnson}(2020)}]{timofeyuk2020}%
  \BibitemOpen
  \bibfield  {author} {\bibinfo {author} {\bibfnamefont {N.}~\bibnamefont
  {Timofeyuk}}\ and\ \bibinfo {author} {\bibfnamefont {R.}~\bibnamefont
  {Johnson}},\ }\href
  {https://doi.org/https://doi.org/10.1016/j.ppnp.2019.103738} {\bibfield
  {journal} {\bibinfo  {journal} {Progress in Particle and Nuclear Physics}\
  }\textbf {\bibinfo {volume} {111}},\ \bibinfo {pages} {103738} (\bibinfo
  {year} {2020})}\BibitemShut {NoStop}%
\bibitem [{\citenamefont {Kamimura}\ \emph {et~al.}(1986)\citenamefont
  {Kamimura}, \citenamefont {Yahiro}, \citenamefont {Iseri}, \citenamefont
  {Sakuragi}, \citenamefont {Kameyama},\ and\ \citenamefont
  {Kawai}}]{Kamimura1986}%
  \BibitemOpen
  \bibfield  {author} {\bibinfo {author} {\bibfnamefont {M.}~\bibnamefont
  {Kamimura}}, \bibinfo {author} {\bibfnamefont {M.}~\bibnamefont {Yahiro}},
  \bibinfo {author} {\bibfnamefont {Y.}~\bibnamefont {Iseri}}, \bibinfo
  {author} {\bibfnamefont {Y.}~\bibnamefont {Sakuragi}}, \bibinfo {author}
  {\bibfnamefont {H.}~\bibnamefont {Kameyama}},\ and\ \bibinfo {author}
  {\bibfnamefont {M.}~\bibnamefont {Kawai}},\ }\href
  {https://doi.org/10.1143/PTPS.89.1} {\bibfield  {journal} {\bibinfo
  {journal} {Progress of Theoretical Physics Supplement}\ }\textbf {\bibinfo
  {volume} {89}},\ \bibinfo {pages} {1} (\bibinfo {year} {1986})}\BibitemShut
  {NoStop}%
\bibitem [{\citenamefont {Austern}\ \emph {et~al.}(1987)\citenamefont
  {Austern}, \citenamefont {Iseri}, \citenamefont {Kamimura}, \citenamefont
  {Kawai}, \citenamefont {Rawitscher},\ and\ \citenamefont
  {Yahiro}}]{austern1987d}%
  \BibitemOpen
  \bibfield  {author} {\bibinfo {author} {\bibfnamefont {N.}~\bibnamefont
  {Austern}}, \bibinfo {author} {\bibfnamefont {Y.}~\bibnamefont {Iseri}},
  \bibinfo {author} {\bibfnamefont {M.}~\bibnamefont {Kamimura}}, \bibinfo
  {author} {\bibfnamefont {M.}~\bibnamefont {Kawai}}, \bibinfo {author}
  {\bibfnamefont {G.}~\bibnamefont {Rawitscher}},\ and\ \bibinfo {author}
  {\bibfnamefont {M.}~\bibnamefont {Yahiro}},\ }\href
  {https://doi.org/https://doi.org/10.1016/0370-1573(87)90094-9} {\bibfield
  {journal} {\bibinfo  {journal} {Physics Reports}\ }\textbf {\bibinfo {volume}
  {154}},\ \bibinfo {pages} {125} (\bibinfo {year} {1987})}\BibitemShut
  {NoStop}%
\bibitem [{\citenamefont {Deltuva}\ and\ \citenamefont
  {Fonseca}(2009)}]{deltuva2009}%
  \BibitemOpen
  \bibfield  {author} {\bibinfo {author} {\bibfnamefont {A.}~\bibnamefont
  {Deltuva}}\ and\ \bibinfo {author} {\bibfnamefont {A.~C.}\ \bibnamefont
  {Fonseca}},\ }\href {https://doi.org/10.1103/PhysRevC.79.014606} {\bibfield
  {journal} {\bibinfo  {journal} {Phys. Rev. C}\ }\textbf {\bibinfo {volume}
  {79}},\ \bibinfo {pages} {014606} (\bibinfo {year} {2009})}\BibitemShut
  {NoStop}%
\bibitem [{\citenamefont {Upadhyay}\ \emph {et~al.}(2012)\citenamefont
  {Upadhyay}, \citenamefont {Deltuva},\ and\ \citenamefont
  {Nunes}}]{upadhyay2012}%
  \BibitemOpen
  \bibfield  {author} {\bibinfo {author} {\bibfnamefont {N.~J.}\ \bibnamefont
  {Upadhyay}}, \bibinfo {author} {\bibfnamefont {A.}~\bibnamefont {Deltuva}},\
  and\ \bibinfo {author} {\bibfnamefont {F.~M.}\ \bibnamefont {Nunes}},\ }\href
  {https://doi.org/10.1103/PhysRevC.85.054621} {\bibfield  {journal} {\bibinfo
  {journal} {Phys. Rev. C}\ }\textbf {\bibinfo {volume} {85}},\ \bibinfo
  {pages} {054621} (\bibinfo {year} {2012})}\BibitemShut {NoStop}%
\bibitem [{\citenamefont {Ogata}\ and\ \citenamefont
  {Yoshida}(2016)}]{ogata2016}%
  \BibitemOpen
  \bibfield  {author} {\bibinfo {author} {\bibfnamefont {K.}~\bibnamefont
  {Ogata}}\ and\ \bibinfo {author} {\bibfnamefont {K.}~\bibnamefont
  {Yoshida}},\ }\href {https://doi.org/10.1103/PhysRevC.94.051603} {\bibfield
  {journal} {\bibinfo  {journal} {Phys. Rev. C}\ }\textbf {\bibinfo {volume}
  {94}},\ \bibinfo {pages} {051603(R)} (\bibinfo {year} {2016})}\BibitemShut
  {NoStop}%
\bibitem [{\citenamefont {Potel}\ \emph {et~al.}(2017)\citenamefont {Potel},
  \citenamefont {Perdikakis}, \citenamefont {Carlson}, \citenamefont
  {Atkinson}, \citenamefont {Dickhoff}, \citenamefont {Escher}, \citenamefont
  {Hussein}, \citenamefont {Lei}, \citenamefont {Li}, \citenamefont
  {Macchiavelli}, \citenamefont {Moro}, \citenamefont {Nunes}, \citenamefont
  {Pain},\ and\ \citenamefont {Rotureau}}]{potel2017}%
  \BibitemOpen
  \bibfield  {author} {\bibinfo {author} {\bibfnamefont {G.}~\bibnamefont
  {Potel}}, \bibinfo {author} {\bibfnamefont {G.}~\bibnamefont {Perdikakis}},
  \bibinfo {author} {\bibfnamefont {B.~V.}\ \bibnamefont {Carlson}}, \bibinfo
  {author} {\bibfnamefont {M.~C.}\ \bibnamefont {Atkinson}}, \bibinfo {author}
  {\bibfnamefont {W.~H.}\ \bibnamefont {Dickhoff}}, \bibinfo {author}
  {\bibfnamefont {J.~E.}\ \bibnamefont {Escher}}, \bibinfo {author}
  {\bibfnamefont {M.~S.}\ \bibnamefont {Hussein}}, \bibinfo {author}
  {\bibfnamefont {J.}~\bibnamefont {Lei}}, \bibinfo {author} {\bibfnamefont
  {W.}~\bibnamefont {Li}}, \bibinfo {author} {\bibfnamefont {A.~O.}\
  \bibnamefont {Macchiavelli}}, \bibinfo {author} {\bibfnamefont {A.~M.}\
  \bibnamefont {Moro}}, \bibinfo {author} {\bibfnamefont {F.~M.}\ \bibnamefont
  {Nunes}}, \bibinfo {author} {\bibfnamefont {S.~D.}\ \bibnamefont {Pain}},\
  and\ \bibinfo {author} {\bibfnamefont {J.}~\bibnamefont {Rotureau}},\ }\href
  {https://doi.org/10.1140/epja/i2017-12371-9} {\bibfield  {journal} {\bibinfo
  {journal} {The European Physical Journal A}\ }\textbf {\bibinfo {volume}
  {53}},\ \bibinfo {pages} {178} (\bibinfo {year} {2017})}\BibitemShut
  {NoStop}%
\bibitem [{\citenamefont {Moeslang}\ \emph {et~al.}(2006)\citenamefont
  {Moeslang}, \citenamefont {Heinzel}, \citenamefont {Matsui},\ and\
  \citenamefont {Sugimoto}}]{Moeslang2006}%
  \BibitemOpen
  \bibfield  {author} {\bibinfo {author} {\bibfnamefont {A.}~\bibnamefont
  {Moeslang}}, \bibinfo {author} {\bibfnamefont {V.}~\bibnamefont {Heinzel}},
  \bibinfo {author} {\bibfnamefont {H.}~\bibnamefont {Matsui}},\ and\ \bibinfo
  {author} {\bibfnamefont {M.}~\bibnamefont {Sugimoto}},\ }\href
  {https://doi.org/https://doi.org/10.1016/j.fusengdes.2005.07.044} {\bibfield
  {journal} {\bibinfo  {journal} {Fusion Engineering and Design}\ }\textbf
  {\bibinfo {volume} {81}},\ \bibinfo {pages} {863} (\bibinfo {year} {2006})},\
  \bibinfo {note} {proceedings of the Seventh International Symposium on Fusion
  Nuclear Technology}\BibitemShut {NoStop}%
\bibitem [{\citenamefont {Nakayama}\ \emph {et~al.}(2016)\citenamefont
  {Nakayama}, \citenamefont {Kouno}, \citenamefont {Watanabe}, \citenamefont
  {Iwamoto},\ and\ \citenamefont {Ogata}}]{nakayama2016}%
  \BibitemOpen
  \bibfield  {author} {\bibinfo {author} {\bibfnamefont {S.}~\bibnamefont
  {Nakayama}}, \bibinfo {author} {\bibfnamefont {H.}~\bibnamefont {Kouno}},
  \bibinfo {author} {\bibfnamefont {Y.}~\bibnamefont {Watanabe}}, \bibinfo
  {author} {\bibfnamefont {O.}~\bibnamefont {Iwamoto}},\ and\ \bibinfo {author}
  {\bibfnamefont {K.}~\bibnamefont {Ogata}},\ }\href
  {https://doi.org/10.1103/PhysRevC.94.014618} {\bibfield  {journal} {\bibinfo
  {journal} {Phys. Rev. C}\ }\textbf {\bibinfo {volume} {94}},\ \bibinfo
  {pages} {014618} (\bibinfo {year} {2016})}\BibitemShut {NoStop}%
\bibitem [{\citenamefont {Nakayama}\ \emph {et~al.}(2020)\citenamefont
  {Nakayama}, \citenamefont {Iwamoto},\ and\ \citenamefont
  {Watanabe}}]{nakayama2020}%
  \BibitemOpen
  \bibfield  {author} {\bibinfo {author} {\bibfnamefont {S.}~\bibnamefont
  {Nakayama}}, \bibinfo {author} {\bibfnamefont {O.}~\bibnamefont {Iwamoto}},\
  and\ \bibinfo {author} {\bibfnamefont {Y.}~\bibnamefont {Watanabe}},\ }\href
  {https://doi.org/10.1051/epjconf/202023903014} {\bibfield  {journal}
  {\bibinfo  {journal} {EPJ Web Conf.}\ }\textbf {\bibinfo {volume} {239}},\
  \bibinfo {pages} {03014} (\bibinfo {year} {2020})}\BibitemShut {NoStop}%
\bibitem [{\citenamefont {Nakayama}\ \emph {et~al.}(2021)\citenamefont
  {Nakayama}, \citenamefont {Iwamoto}, \citenamefont {Watanabe},\ and\
  \citenamefont {Ogata}}]{nakayama2021}%
  \BibitemOpen
  \bibfield  {author} {\bibinfo {author} {\bibfnamefont {S.}~\bibnamefont
  {Nakayama}}, \bibinfo {author} {\bibfnamefont {O.}~\bibnamefont {Iwamoto}},
  \bibinfo {author} {\bibfnamefont {Y.}~\bibnamefont {Watanabe}},\ and\
  \bibinfo {author} {\bibfnamefont {K.}~\bibnamefont {Ogata}},\ }\href
  {https://doi.org/10.1080/00223131.2020.1870010} {\bibfield  {journal}
  {\bibinfo  {journal} {Journal of Nuclear Science and Technology}\ }\textbf
  {\bibinfo {volume} {58}},\ \bibinfo {pages} {805} (\bibinfo {year}
  {2021})}\BibitemShut {NoStop}%
\bibitem [{\citenamefont {Glauber}(1959)}]{glauber1959}%
  \BibitemOpen
  \bibfield  {author} {\bibinfo {author} {\bibfnamefont {R.~J.}\ \bibnamefont
  {Glauber}},\ }\href@noop {} {\bibfield  {journal} {\bibinfo  {journal} {in
  Lectures in Theoretical Physics,Vol.}\ }\textbf {\bibinfo {volume} {1}},\
  \bibinfo {pages} {p. 315} (\bibinfo {year} {Interscience, New York,
  1959})}\BibitemShut {NoStop}%
\bibitem [{\citenamefont {Hussein}\ and\ \citenamefont
  {McVoy}(1985)}]{Hussein1985}%
  \BibitemOpen
  \bibfield  {author} {\bibinfo {author} {\bibfnamefont {M.}~\bibnamefont
  {Hussein}}\ and\ \bibinfo {author} {\bibfnamefont {K.}~\bibnamefont
  {McVoy}},\ }\href
  {https://doi.org/https://doi.org/10.1016/0375-9474(85)90364-1} {\bibfield
  {journal} {\bibinfo  {journal} {Nuclear Physics A}\ }\textbf {\bibinfo
  {volume} {445}},\ \bibinfo {pages} {124} (\bibinfo {year}
  {1985})}\BibitemShut {NoStop}%
\bibitem [{\citenamefont {Hencken}\ \emph {et~al.}(1996)\citenamefont
  {Hencken}, \citenamefont {Bertsch},\ and\ \citenamefont
  {Esbensen}}]{hencken1996a}%
  \BibitemOpen
  \bibfield  {author} {\bibinfo {author} {\bibfnamefont {K.}~\bibnamefont
  {Hencken}}, \bibinfo {author} {\bibfnamefont {G.}~\bibnamefont {Bertsch}},\
  and\ \bibinfo {author} {\bibfnamefont {H.}~\bibnamefont {Esbensen}},\ }\href
  {https://doi.org/10.1103/PhysRevC.54.3043} {\bibfield  {journal} {\bibinfo
  {journal} {Phys. Rev. C}\ }\textbf {\bibinfo {volume} {54}},\ \bibinfo
  {pages} {3043} (\bibinfo {year} {1996})}\BibitemShut {NoStop}%
\bibitem [{\citenamefont {Ichimura}\ \emph {et~al.}(1985)\citenamefont
  {Ichimura}, \citenamefont {Austern},\ and\ \citenamefont
  {Vincent}}]{ichimura1985}%
  \BibitemOpen
  \bibfield  {author} {\bibinfo {author} {\bibfnamefont {M.}~\bibnamefont
  {Ichimura}}, \bibinfo {author} {\bibfnamefont {N.}~\bibnamefont {Austern}},\
  and\ \bibinfo {author} {\bibfnamefont {C.~M.}\ \bibnamefont {Vincent}},\
  }\href {https://doi.org/10.1103/PhysRevC.32.431} {\bibfield  {journal}
  {\bibinfo  {journal} {Phys. Rev. C}\ }\textbf {\bibinfo {volume} {32}},\
  \bibinfo {pages} {431} (\bibinfo {year} {1985})}\BibitemShut {NoStop}%
\bibitem [{\citenamefont {Lei}\ and\ \citenamefont
  {Moro}(2015{\natexlab{a}})}]{lei2015}%
  \BibitemOpen
  \bibfield  {author} {\bibinfo {author} {\bibfnamefont {J.}~\bibnamefont
  {Lei}}\ and\ \bibinfo {author} {\bibfnamefont {A.~M.}\ \bibnamefont {Moro}},\
  }\href {https://doi.org/10.1103/PhysRevC.92.044616} {\bibfield  {journal}
  {\bibinfo  {journal} {Phys. Rev. C}\ }\textbf {\bibinfo {volume} {92}},\
  \bibinfo {pages} {044616} (\bibinfo {year} {2015}{\natexlab{a}})}\BibitemShut
  {NoStop}%
\bibitem [{\citenamefont {Lei}\ and\ \citenamefont
  {Moro}(2015{\natexlab{b}})}]{lei2015a}%
  \BibitemOpen
  \bibfield  {author} {\bibinfo {author} {\bibfnamefont {J.}~\bibnamefont
  {Lei}}\ and\ \bibinfo {author} {\bibfnamefont {A.~M.}\ \bibnamefont {Moro}},\
  }\href {https://doi.org/10.1103/PhysRevC.92.061602} {\bibfield  {journal}
  {\bibinfo  {journal} {Phys. Rev. C}\ }\textbf {\bibinfo {volume} {92}},\
  \bibinfo {pages} {061602(R)} (\bibinfo {year}
  {2015}{\natexlab{b}})}\BibitemShut {NoStop}%
\bibitem [{\citenamefont {Feshbach}\ \emph {et~al.}(1980)\citenamefont
  {Feshbach}, \citenamefont {Kerman},\ and\ \citenamefont
  {Koonin}}]{Feshbach1980}%
  \BibitemOpen
  \bibfield  {author} {\bibinfo {author} {\bibfnamefont {H.}~\bibnamefont
  {Feshbach}}, \bibinfo {author} {\bibfnamefont {A.}~\bibnamefont {Kerman}},\
  and\ \bibinfo {author} {\bibfnamefont {S.}~\bibnamefont {Koonin}},\ }\href
  {https://doi.org/https://doi.org/10.1016/0003-4916(80)90140-2} {\bibfield
  {journal} {\bibinfo  {journal} {Annals of Physics}\ }\textbf {\bibinfo
  {volume} {125}},\ \bibinfo {pages} {429} (\bibinfo {year}
  {1980})}\BibitemShut {NoStop}%
\bibitem [{\citenamefont {Nishioka}\ \emph {et~al.}(1988)\citenamefont
  {Nishioka}, \citenamefont {Weidenmüller},\ and\ \citenamefont
  {Yoshida}}]{nishioka1988}%
  \BibitemOpen
  \bibfield  {author} {\bibinfo {author} {\bibfnamefont {H.}~\bibnamefont
  {Nishioka}}, \bibinfo {author} {\bibfnamefont {H.}~\bibnamefont
  {Weidenmüller}},\ and\ \bibinfo {author} {\bibfnamefont {S.}~\bibnamefont
  {Yoshida}},\ }\href
  {https://doi.org/https://doi.org/10.1016/0003-4916(88)90250-3} {\bibfield
  {journal} {\bibinfo  {journal} {Annals of Physics}\ }\textbf {\bibinfo
  {volume} {183}},\ \bibinfo {pages} {166} (\bibinfo {year}
  {1988})}\BibitemShut {NoStop}%
\bibitem [{\citenamefont {Tamura}\ \emph {et~al.}(1982)\citenamefont {Tamura},
  \citenamefont {Udagawa},\ and\ \citenamefont {Lenske}}]{tamura1982}%
  \BibitemOpen
  \bibfield  {author} {\bibinfo {author} {\bibfnamefont {T.}~\bibnamefont
  {Tamura}}, \bibinfo {author} {\bibfnamefont {T.}~\bibnamefont {Udagawa}},\
  and\ \bibinfo {author} {\bibfnamefont {H.}~\bibnamefont {Lenske}},\ }\href
  {https://doi.org/10.1103/PhysRevC.26.379} {\bibfield  {journal} {\bibinfo
  {journal} {Phys. Rev. C}\ }\textbf {\bibinfo {volume} {26}},\ \bibinfo
  {pages} {379} (\bibinfo {year} {1982})}\BibitemShut {NoStop}%
\bibitem [{\citenamefont {Luo}\ and\ \citenamefont {Kawai}(1991)}]{Luo1991}%
  \BibitemOpen
  \bibfield  {author} {\bibinfo {author} {\bibfnamefont {Y.~L.}\ \bibnamefont
  {Luo}}\ and\ \bibinfo {author} {\bibfnamefont {M.}~\bibnamefont {Kawai}},\
  }\href {https://doi.org/10.1103/PhysRevC.43.2367} {\bibfield  {journal}
  {\bibinfo  {journal} {Phys. Rev. C}\ }\textbf {\bibinfo {volume} {43}},\
  \bibinfo {pages} {2367} (\bibinfo {year} {1991})}\BibitemShut {NoStop}%
\bibitem [{\citenamefont {Kawai}\ and\ \citenamefont
  {Weidenm\"uller}(1992)}]{kawai1992}%
  \BibitemOpen
  \bibfield  {author} {\bibinfo {author} {\bibfnamefont {M.}~\bibnamefont
  {Kawai}}\ and\ \bibinfo {author} {\bibfnamefont {H.~A.}\ \bibnamefont
  {Weidenm\"uller}},\ }\href {https://doi.org/10.1103/PhysRevC.45.1856}
  {\bibfield  {journal} {\bibinfo  {journal} {Phys. Rev. C}\ }\textbf {\bibinfo
  {volume} {45}},\ \bibinfo {pages} {1856} (\bibinfo {year}
  {1992})}\BibitemShut {NoStop}%
\bibitem [{\citenamefont {Watanabe}\ \emph {et~al.}(1999)\citenamefont
  {Watanabe}, \citenamefont {Kuwata}, \citenamefont {Weili}, \citenamefont
  {Higashi}, \citenamefont {Shinohara}, \citenamefont {Kohno}, \citenamefont
  {Ogata},\ and\ \citenamefont {Kawai}}]{watanabe1999}%
  \BibitemOpen
  \bibfield  {author} {\bibinfo {author} {\bibfnamefont {Y.}~\bibnamefont
  {Watanabe}}, \bibinfo {author} {\bibfnamefont {R.}~\bibnamefont {Kuwata}},
  \bibinfo {author} {\bibfnamefont {S.}~\bibnamefont {Weili}}, \bibinfo
  {author} {\bibfnamefont {M.}~\bibnamefont {Higashi}}, \bibinfo {author}
  {\bibfnamefont {H.}~\bibnamefont {Shinohara}}, \bibinfo {author}
  {\bibfnamefont {M.}~\bibnamefont {Kohno}}, \bibinfo {author} {\bibfnamefont
  {K.}~\bibnamefont {Ogata}},\ and\ \bibinfo {author} {\bibfnamefont
  {M.}~\bibnamefont {Kawai}},\ }\href
  {https://doi.org/10.1103/PhysRevC.59.2136} {\bibfield  {journal} {\bibinfo
  {journal} {Phys. Rev. C}\ }\textbf {\bibinfo {volume} {59}},\ \bibinfo
  {pages} {2136} (\bibinfo {year} {1999})}\BibitemShut {NoStop}%
\bibitem [{\citenamefont {Ogata}\ \emph {et~al.}(1999)\citenamefont {Ogata},
  \citenamefont {Kawai}, \citenamefont {Watanabe}, \citenamefont {Weili},\ and\
  \citenamefont {Kohno}}]{ogata1999}%
  \BibitemOpen
  \bibfield  {author} {\bibinfo {author} {\bibfnamefont {K.}~\bibnamefont
  {Ogata}}, \bibinfo {author} {\bibfnamefont {M.}~\bibnamefont {Kawai}},
  \bibinfo {author} {\bibfnamefont {Y.}~\bibnamefont {Watanabe}}, \bibinfo
  {author} {\bibfnamefont {S.}~\bibnamefont {Weili}},\ and\ \bibinfo {author}
  {\bibfnamefont {M.}~\bibnamefont {Kohno}},\ }\href
  {https://doi.org/10.1103/PhysRevC.60.054605} {\bibfield  {journal} {\bibinfo
  {journal} {Phys. Rev. C}\ }\textbf {\bibinfo {volume} {60}},\ \bibinfo
  {pages} {054605} (\bibinfo {year} {1999})}\BibitemShut {NoStop}%
\bibitem [{\citenamefont {Weili}\ \emph {et~al.}(1999)\citenamefont {Weili},
  \citenamefont {Watanabe}, \citenamefont {Kohno}, \citenamefont {Ogata},\ and\
  \citenamefont {Kawai}}]{weili1999}%
  \BibitemOpen
  \bibfield  {author} {\bibinfo {author} {\bibfnamefont {S.}~\bibnamefont
  {Weili}}, \bibinfo {author} {\bibfnamefont {Y.}~\bibnamefont {Watanabe}},
  \bibinfo {author} {\bibfnamefont {M.}~\bibnamefont {Kohno}}, \bibinfo
  {author} {\bibfnamefont {K.}~\bibnamefont {Ogata}},\ and\ \bibinfo {author}
  {\bibfnamefont {M.}~\bibnamefont {Kawai}},\ }\href
  {https://doi.org/10.1103/PhysRevC.60.064605} {\bibfield  {journal} {\bibinfo
  {journal} {Phys. Rev. C}\ }\textbf {\bibinfo {volume} {60}},\ \bibinfo
  {pages} {064605} (\bibinfo {year} {1999})}\BibitemShut {NoStop}%
\bibitem [{\citenamefont {Ogata}\ \emph {et~al.}(2002)\citenamefont {Ogata},
  \citenamefont {Watanabe}, \citenamefont {Weili}, \citenamefont {Kohno},\ and\
  \citenamefont {Kawai}}]{ogata2002}%
  \BibitemOpen
  \bibfield  {author} {\bibinfo {author} {\bibfnamefont {K.}~\bibnamefont
  {Ogata}}, \bibinfo {author} {\bibfnamefont {Y.}~\bibnamefont {Watanabe}},
  \bibinfo {author} {\bibfnamefont {S.}~\bibnamefont {Weili}}, \bibinfo
  {author} {\bibfnamefont {M.}~\bibnamefont {Kohno}},\ and\ \bibinfo {author}
  {\bibfnamefont {M.}~\bibnamefont {Kawai}},\ }\href
  {https://doi.org/https://doi.org/10.1016/S0375-9474(01)01339-2} {\bibfield
  {journal} {\bibinfo  {journal} {Nuclear Physics A}\ }\textbf {\bibinfo
  {volume} {703}},\ \bibinfo {pages} {152} (\bibinfo {year}
  {2002})}\BibitemShut {NoStop}%
\bibitem [{\citenamefont {Watanabe}\ and\ \citenamefont
  {Kawai}(1993)}]{watanabe1993}%
  \BibitemOpen
  \bibfield  {author} {\bibinfo {author} {\bibfnamefont {Y.}~\bibnamefont
  {Watanabe}}\ and\ \bibinfo {author} {\bibfnamefont {M.}~\bibnamefont
  {Kawai}},\ }\href
  {https://doi.org/https://doi.org/10.1016/0375-9474(93)90082-9} {\bibfield
  {journal} {\bibinfo  {journal} {Nuclear Physics A}\ }\textbf {\bibinfo
  {volume} {560}},\ \bibinfo {pages} {43} (\bibinfo {year} {1993})}\BibitemShut
  {NoStop}%
\bibitem [{\citenamefont {Förtsch}\ \emph {et~al.}(2002)\citenamefont
  {Förtsch}, \citenamefont {Ridikas}, \citenamefont {Mittig}, \citenamefont
  {Savajols}, \citenamefont {Roussel–Chomaz}, \citenamefont {Steyn},\ and\
  \citenamefont {Lawrie}}]{Fortsch2002}%
  \BibitemOpen
  \bibfield  {author} {\bibinfo {author} {\bibfnamefont {S.}~\bibnamefont
  {Förtsch}}, \bibinfo {author} {\bibfnamefont {D.}~\bibnamefont {Ridikas}},
  \bibinfo {author} {\bibfnamefont {W.}~\bibnamefont {Mittig}}, \bibinfo
  {author} {\bibfnamefont {H.}~\bibnamefont {Savajols}}, \bibinfo {author}
  {\bibfnamefont {P.}~\bibnamefont {Roussel–Chomaz}}, \bibinfo {author}
  {\bibfnamefont {G.}~\bibnamefont {Steyn}},\ and\ \bibinfo {author}
  {\bibfnamefont {J.}~\bibnamefont {Lawrie}},\ }\href
  {https://doi.org/10.1080/00223131.2002.10875217} {\bibfield  {journal}
  {\bibinfo  {journal} {Journal of Nuclear Science and Technology}\ }\textbf
  {\bibinfo {volume} {39}},\ \bibinfo {pages} {792} (\bibinfo {year}
  {2002})}\BibitemShut {NoStop}%
\bibitem [{\citenamefont {Wu}\ \emph {et~al.}(1979)\citenamefont {Wu},
  \citenamefont {Chang},\ and\ \citenamefont {Holmgren}}]{Wu1979}%
  \BibitemOpen
  \bibfield  {author} {\bibinfo {author} {\bibfnamefont {J.~R.}\ \bibnamefont
  {Wu}}, \bibinfo {author} {\bibfnamefont {C.~C.}\ \bibnamefont {Chang}},\ and\
  \bibinfo {author} {\bibfnamefont {H.~D.}\ \bibnamefont {Holmgren}},\ }\href
  {https://doi.org/10.1103/PhysRevC.19.370} {\bibfield  {journal} {\bibinfo
  {journal} {Phys. Rev. C}\ }\textbf {\bibinfo {volume} {19}},\ \bibinfo
  {pages} {370} (\bibinfo {year} {1979})}\BibitemShut {NoStop}%
\bibitem [{\citenamefont {An}\ and\ \citenamefont {Cai}(2006)}]{An2006}%
  \BibitemOpen
  \bibfield  {author} {\bibinfo {author} {\bibfnamefont {H.}~\bibnamefont
  {An}}\ and\ \bibinfo {author} {\bibfnamefont {C.}~\bibnamefont {Cai}},\
  }\href {https://doi.org/10.1103/PhysRevC.73.054605} {\bibfield  {journal}
  {\bibinfo  {journal} {Phys. Rev. C}\ }\textbf {\bibinfo {volume} {73}},\
  \bibinfo {pages} {054605} (\bibinfo {year} {2006})}\BibitemShut {NoStop}%
\bibitem [{\citenamefont {Perey}\ and\ \citenamefont {Buck}(1962)}]{Perey1962}%
  \BibitemOpen
  \bibfield  {author} {\bibinfo {author} {\bibfnamefont {F.}~\bibnamefont
  {Perey}}\ and\ \bibinfo {author} {\bibfnamefont {B.}~\bibnamefont {Buck}},\
  }\href {https://doi.org/https://doi.org/10.1016/0029-5582(62)90345-0}
  {\bibfield  {journal} {\bibinfo  {journal} {Nuclear Physics}\ }\textbf
  {\bibinfo {volume} {32}},\ \bibinfo {pages} {353} (\bibinfo {year}
  {1962})}\BibitemShut {NoStop}%
\bibitem [{\citenamefont {TWOFNR}()}]{TWOFNR}%
  \BibitemOpen
  \bibfield  {author} {\bibinfo {author} {\bibnamefont {TWOFNR}},\ }\href@noop
  {} {\bibinfo {title} {Usermanual}},\ \bibinfo {note}
  {{https://people.nscl.msu.edu/
  ~brown/reaction-codes/twofnr/twofnr.pdf.}}\BibitemShut {Stop}%
\bibitem [{\citenamefont {Chazono}\ \emph {et~al.}(2022)\citenamefont
  {Chazono}, \citenamefont {Yoshida},\ and\ \citenamefont
  {Ogata}}]{chazono2022a}%
  \BibitemOpen
  \bibfield  {author} {\bibinfo {author} {\bibfnamefont {Y.}~\bibnamefont
  {Chazono}}, \bibinfo {author} {\bibfnamefont {K.}~\bibnamefont {Yoshida}},\
  and\ \bibinfo {author} {\bibfnamefont {K.}~\bibnamefont {Ogata}},\ }\href
  {https://doi.org/10.1103/PhysRevC.106.064613} {\bibfield  {journal} {\bibinfo
   {journal} {Phys. Rev. C}\ }\textbf {\bibinfo {volume} {106}},\ \bibinfo
  {pages} {064613} (\bibinfo {year} {2022})}\BibitemShut {NoStop}%
\bibitem [{\citenamefont {Love}\ and\ \citenamefont {Franey}(1981)}]{Love1981}%
  \BibitemOpen
  \bibfield  {author} {\bibinfo {author} {\bibfnamefont {W.~G.}\ \bibnamefont
  {Love}}\ and\ \bibinfo {author} {\bibfnamefont {M.~A.}\ \bibnamefont
  {Franey}},\ }\href {https://doi.org/10.1103/PhysRevC.24.1073} {\bibfield
  {journal} {\bibinfo  {journal} {Phys. Rev. C}\ }\textbf {\bibinfo {volume}
  {24}},\ \bibinfo {pages} {1073} (\bibinfo {year} {1981})}\BibitemShut
  {NoStop}%
\bibitem [{\citenamefont {Franey}\ and\ \citenamefont
  {Love}(1985)}]{Franey1985}%
  \BibitemOpen
  \bibfield  {author} {\bibinfo {author} {\bibfnamefont {M.~A.}\ \bibnamefont
  {Franey}}\ and\ \bibinfo {author} {\bibfnamefont {W.~G.}\ \bibnamefont
  {Love}},\ }\href {https://doi.org/10.1103/PhysRevC.31.488} {\bibfield
  {journal} {\bibinfo  {journal} {Phys. Rev. C}\ }\textbf {\bibinfo {volume}
  {31}},\ \bibinfo {pages} {488} (\bibinfo {year} {1985})}\BibitemShut
  {NoStop}%
\bibitem [{\citenamefont {Minomo}\ \emph {et~al.}(2010)\citenamefont {Minomo},
  \citenamefont {Ogata}, \citenamefont {Kohno}, \citenamefont {Shimizu},\ and\
  \citenamefont {Yahiro}}]{Minomo2010}%
  \BibitemOpen
  \bibfield  {author} {\bibinfo {author} {\bibfnamefont {K.}~\bibnamefont
  {Minomo}}, \bibinfo {author} {\bibfnamefont {K.}~\bibnamefont {Ogata}},
  \bibinfo {author} {\bibfnamefont {M.}~\bibnamefont {Kohno}}, \bibinfo
  {author} {\bibfnamefont {Y.~R.}\ \bibnamefont {Shimizu}},\ and\ \bibinfo
  {author} {\bibfnamefont {M.}~\bibnamefont {Yahiro}},\ }\href
  {https://doi.org/10.1088/0954-3899/37/8/085011} {\bibfield  {journal}
  {\bibinfo  {journal} {Journal of Physics G: Nuclear and Particle Physics}\
  }\textbf {\bibinfo {volume} {37}},\ \bibinfo {pages} {085011} (\bibinfo
  {year} {2010})}\BibitemShut {NoStop}%
\bibitem [{\citenamefont {Yoshida}\ \emph {et~al.}(2016)\citenamefont
  {Yoshida}, \citenamefont {Minomo},\ and\ \citenamefont
  {Ogata}}]{yoshida2016}%
  \BibitemOpen
  \bibfield  {author} {\bibinfo {author} {\bibfnamefont {K.}~\bibnamefont
  {Yoshida}}, \bibinfo {author} {\bibfnamefont {K.}~\bibnamefont {Minomo}},\
  and\ \bibinfo {author} {\bibfnamefont {K.}~\bibnamefont {Ogata}},\ }\href
  {https://doi.org/10.1103/PhysRevC.94.044604} {\bibfield  {journal} {\bibinfo
  {journal} {Phys. Rev. C}\ }\textbf {\bibinfo {volume} {94}},\ \bibinfo
  {pages} {044604} (\bibinfo {year} {2016})}\BibitemShut {NoStop}%
\end{thebibliography}%

\end{document}